\documentstyle[12pt]{article}
\renewcommand{\baselinestretch}{1.65}

\newfont{\Mb}{msbm10}

\begin{document}

\renewcommand{\baselinestretch}{1.1}
\title{\Large \bf Dirac's \AE ther in Curved Spacetime-II: \\ The {\it Geometric Amplification} of the Cosmic Magnetic Induction }
\author{
Marcelo Carvalho \thanks{ {\sc e-mail:
marcelo\underline
{\hspace{2mm}}carv@hotmail.com} } \\
\normalsize{Waseda University} \\
\normalsize{Department of Mathematics }\\
\normalsize{3-4-1 Okubo, Shinjuku-ku, Tokyo 169, Japan}
\\
\\
A.L. Oliveira \thanks{ {\sc internet:
alexandr@ov.ufrj.br} } $~^{\sharp}~$$^{\S}$\\
C.R. Raba\c{c}a \thanks{ {\sc internet:
rabaca@nova.ov.ufrj.br} } $^{\S}$ \\
\normalsize{Universidade Federal do Rio de Janeiro} \\
\normalsize{Observat\'{o}rio do Valongo }\\
\normalsize{20080-090 Rio de Janeiro -- RJ, Brazil}
\\
\\  
$~^{\sharp}~$\normalsize{Grupo de F\'{\i}sica e Astrof\'{\i}sica Relativista (GFAR)}  \\
$^{\S}~$\normalsize{Grupo de Estudos Avan\c{c}ados e Modelagem em Astrof\'{\i}sica (GMAC)}
 \vspace{3mm}
\\
   }
\date{}
\maketitle 
\vspace{2mm}
\begin{abstract} \vspace{2mm}
We search for an amplification mechanism of the seed cosmic
magnetic induction by studying a new version of the Dirac's \ae
ther in a curved cosmological background. We find that an
amplification takes place if the scale factor $R(t)$ varies with
the cosmic time, which brings to the magnetic field the effect of
a {\it geometric amplification}.
\end{abstract}
\vspace{2mm}
{\raggedright
\section{Introduction} }  \label{I}
\setcounter{equation}{0}
Information from our universe comes vastly from the propagation of
light throughout the cosmic medium. However, until now the origin
of the cosmic magnetic induction (CMI) in astronomical objects
remains unknown (see \cite{Grasso} for a
comprehensive review). In fact, no theory has completely succeeded
in explaining the evolution of the CMI, from its generation in the
early universe to the present values observed in a multitude of
cosmic scales. In stars, magnetic fields range from $10^8$ T in
the interiors of neutron stars to values of $\approx 1$ T in
sunspots, and down to $10^{-7}$ T in protostellar objects. It is
believed that in stellar interiors the standard dynamo action, in
combination with convective motions and the reconnection of the
field lines, is able to explain the range of values observed. In
galaxies, however, magnetic dynamos leave many unanswered
questions \cite{Parker}, the most immediate of them is the rather
long characteristic time-scales over which they could operate. For
instance, the interstellar medium in our Galaxy has an ordered
field of about $2 \times 10^{-10}$ T, superposed upon which there
is a random component of about 1-2 times this value \cite{Taylor}. Considering that the period of rotation of the
interstellar gas about the Galactic center is $\approx 2.5 \times
10^8$ years, there would have been at most 50 complete rotations
of this gas about the center. The differential rotation of the
ionized gas in the interstellar medium results in the stretching
and amplification of the magnetic field in the disk. Therefore,
any primordial magnetic field would be tightly wound up. The
problem is that this mechanism is not enough to produce an ordered
field. The winding up of the field lines would result in tightly
wound tubes of magnetic flux running in opposite directions. It is
then necessary to have a mechanism that is able to reconnect the
lines of force in order to create the large-scale uniform field.

As we proceed to even larger structures, evidence for large scale
magnetic fields are seen. In the intracluster medium in cluster of
galaxies, fields of $\approx 10^{-10}$ T have been derived from
the diffuse synchrotron radio emission observed from a number of
clusters of galaxies, as well as from the observation of
depolarization of the emission of extended radio sources by the
surrounding intracluster medium. In the intergalactic medium
between clusters of galaxies, we are only able to set an upper
limit for the magnetic field of $10^{-13}$ T, a value provided
from the lack of polarization of the emission reaching us from
distant radio sources. However, it is widely believed that
magnetic inductions are present in the universe whatever the scale
we look for, and new technological developments are in their way
to confirm this (e.g., the SIRTF infrared satellite and the SOAR optical telescope). Whether present-day field values
were built up when the first galaxies formed remains questionable,
but even if this is the case, the process requires the existence
of a very week seed field that was slowly amplified over cosmic
time.

It has been shown that it is possible to generate a seed CMI in a
plasma with no fields present at the recombination time supposing
that there are only variations in the pressure of the electrons in
the plasma. This effect is known as the Biermann battery (see \cite{Parker}). The flow of electrons to lower pressure regions
results in a charged unbalanced plasma, which produces an electric
field opposing the flow of electrons. As a result, the flow stops
and an eletric and magnetic field (emf) is created. This emf
cannot drive a current though, since the integral around any
closed loop in the case of a linear gradient is zero. If, however,
there are variations in the electron density throughout the
plasma, different emfs can be induced in different regions and
then currents flow in the plasma creating a magnetic field.
Nevertheless, this process saturates at about $10^{-25}$ T, since
it is limited by the self-inductance of the current loop itself.
How then to reconcile this extremely low value with the upper
limit found today for the intergalactic medium between clusters?

Recently, one special mechanism has been studied in which the
amplification of the seed CMI is understood as being caused by the
expansion of the cosmological background \cite{Oliveira 2001}. This is
called {\it geometric amplification} because the only agent
responsible is the geometric scale factor $R(t)$. One of the
advantages of this approach is that it does not rule out other
models, while keeping the amplification factor obtained for the
seed CMI within the observational constraint.

The Dirac's \ae ther is a kind of cosmic conducting medium with a
very small conductivity that does not violate the experimental
limits which confirms the maxwellian theory in terrestrial
laboratories. In the first paper \cite{OT}, we
studied the equations of the Dirac's \ae ther coupled to the Proca
field in the background of an Einstein static universe. Here we
search for a {\it geometric amplification} of the seed CMI by
applying to the cosmic medium a recent version of the Dirac's \ae
ther model \cite{Oliveira 2000, Carvalho e Oliveira 2002}. This new
version of the Dirac's \ae ther maintains the most important
features of the original Dirac's model with aditional advantages,
e.g., a new interpretation of the 4-velocity as the velocity of
the different parts of the \ae ther relative  to a generic
observer, inertial or not.

This work is organized as follows. In the next section, we
describe the model, the symmetries of the proposed field, and
solve the equations numerically. In the last section we devote
ourselves to the discussion and concluding remarks.
\vspace{1cm}	
{\raggedright
\section{Model and Results} }  \label{MR}
\setcounter{equation}{0}
In this paper we use the new equations of the Dirac's \ae ther
coupled with geometry. For a generic observer (inertial or not)
these equations are
\begin{eqnarray}\label{EQPV}
F^{\mu \nu}_{\; ; \mu} + \frac {\sigma} {c} (A^{\mu} v^{\nu} -
A^{\nu} v^{\mu})_{\; ; \mu}  =  J^{\nu}
\label{equ}
\end{eqnarray}
\noindent where the semicolon denotes the covariant derivative.
$A^{\mu}$ is the electromagnetic 4-potential; $ F_{\mu \nu} \equiv
\partial_{\mu}A_{\nu} - \partial_{\nu} A_{\mu}$ are the components
of the electromagnetic field tensor; $v$ is the 4-velocity of the
\ae ther relative to the observer; $c$ is the speed of light; and
$J^{\nu} \equiv (-\sigma /c)v_{\mu}F^{\mu\nu}$. It is interesting
to remember that in a flat space-time an inertial observer moving
with the cosmic \ae ther has velocity components $v^{\mu}_{\ae
ther} = (1;0,0,0)$. Others inertial observers have $v^{\mu} =
\Lambda ^{\mu}_{\nu}\,v^{\nu}_{\ae ther} = \Lambda^{\nu}_{0}$,
with the Lorentz transformation $\Lambda$ relating the observer
and the \ae ther frame. These equations (1) differ from
the ones of our previous model by the presence of a
skew-symmetric term $(A^{\mu} v^{\nu} -  A^{\nu} v^{\mu})_{\; ;
\mu}$ instead of the term $(1 / \lambda ^2){A}^{\nu}$ (that comes
from the Proca term). We adopt a cylindrical coordinate system
$x^ {\mu } = (t;\, \rho ,\,  \phi ,\,  \zeta )$ in a Friedmann
cosmological background. The metric tensor in all three Friedmann
geometries and the field $A^{\mu}$, with cylindrical symmetry, are given by
\begin{eqnarray} \label{ds2-d}
g_{\mu \nu} &=&  diag\left[R^{2}(t)(1\,; -1 \,, -u^2(\rho) \, , - w^2(\rho) )\right], \label{Amu}\\
A^{\mu}  &=&  (0;\,  0,\,  1,\,  0)f(t)/R^2(t),
\end{eqnarray}
\noindent where $u(\rho)$ and $w(\rho)$ are functions of the
coordinate $\rho$. Depending on the geometry of the space-time,
they will define the type of three-geometry (with constant
curvature $k_{c}$) under consideration. $f(t)$ is a function to be
determined by the field equations. The velocity of observers
moving with the cosmic \ae ther is $v_{\mu} = R(t)\,\delta
^{0}_\mu $.

The field strength $F_{\mu \nu}$ has non-zero independent
components $F_{02}$ and $F_{12}$. In an orthonormal basis, the
non-null components of the fields ${\bf E}$ and $c{\bf B}$ are
\begin{eqnarray} \label{Ffis}
E_\phi = - \dot{f} u^2/R^2 , \hspace{1cm} cB_\zeta = 2 f u\;u'/R^2 ,
\end{eqnarray}
\noindent where the dot means $d/dt$, the prime is $d/d\rho$, and
$c$ is the light velocity. The electric field and the magnetic
induction are orthogonal and non-homogeneous, and both depend on
$t$ and $\rho$. Their moduli are
\begin{eqnarray} \label{EeB}
|{\bf E}|= |u \dot{f}|/R^2 \;, \hspace{2cm} |{\bf B}|= 2 |f u'|/(cR^2).
\end{eqnarray}
\noindent Table 1 shows the curvature $k_{c}$ of each one of the
Friedmann's geometries considered, $R(t)$, $u(\rho)$, $w(\rho)$,
and the equation for $f(t)$ that needs to be solved. We are
particularly interested in finding solutions for the CMI that
provide an explanation for the amplification of the magnetic
induction below the the suggested limit of $10^{-13}$ T for the
intergalactic medium between cluster of galaxies.
\begin{table}[h]
\caption{Curvature $k_{c}$, $R(t)$, $u(\rho)$, $w(\rho)$, and equations to be solved. } \label{tabela1}
\begin{center}
\begin{tabular}{c c c c c c} \hline
  $k_{c}$ & $R(t)$ & $u(\rho)$ & $w(\rho)$ & Equation for $f(t)$
\\ \hline \hline
$ 0$ & $(\alpha/2) t^{2}$ & $\rho$ & $1$ & $\ddot{f} -
\alpha t \;(\sigma/c) \;f = 0 $
\\ \hline
$+1$ & $\alpha (1 - \cos t)$ & $\sin\rho$ & $\cos\rho$ &
$\ddot{f} + [4 - \;(\sigma/c)\; \alpha \,\sin t] \;f = 0$
\\ \hline
$-1$ & $\alpha (\cosh t -1) $ &  $\sinh\rho$ & $\cosh\rho$
& $\ddot{f} - [4 + \;(\sigma/c) \; \alpha \, \sinh t] \;f = 0$
\\ \hline
\end{tabular}
\end{center}
\end{table}

Let us now integrate numerically the equations in Table 1, from
the initial cosmic conformal time $t_{i} = 0.0890$ to the final
time $t_{f} = 1.6100$. In standard cosmology these values
correspond respectively to the final stage of the matter-radiation
coupling and our current epoch. $20,000$ integration steps are
performed. We assume that the conductivity of the Dirac's \ae ther
is $\approx 10^{-19}$/s, $\alpha$ is $10^{26}$ m, and the initial
CMI is $\approx 10^{-25}$ T. Our model also includes a weak
initial electric field of magnitude $\approx 10^{-4}$ V/m to be
dissipated during the time evolution. These limits are fixed in
order to provide a realistic value for the modulus of the CMI that
agrees with the one established by the usual theory of the cosmic
fields. These initial values do not perturb the gravitational
field; from a simple calculation it is evident that the
energy-momentum tensor of the electromagnetic and gravitational
fields are related by a factor above $10^{10}$.

In Table 2, we display a small ensemble of points that gives us a
qualitative view of the amplification phenomenon in terms of the
quantities ${\cal E}(t)$ $\equiv$ $|{\bf E}/u| = | \dot{f}|/R^2$ and ${\cal B}(t)$ $\equiv$ $|{\bf B}/u'| = 2 |f |/(cR^2)$, that
for simplicity we will also refer as the electric field and
magnetic induction.

\begin{table}[h]
\caption{Some Results of the Numerical Integration}
\label{tabela3}
\begin{center}
\begin{tabular}{c|c|c|c|c|c|c}
\hline
  & \multicolumn{2}{c|}{Flat ($k_c=0$)} & \multicolumn{2}{c|}{Elliptic ($k_c=+1$)} & \multicolumn{2}{c}{Hyperbolic ($k_c=-1$)} \\
\cline{2-7}
t & $\;log|\,{\cal E}\,|\;$ & $\;log|\,c{\cal B}\,|\;$ &
    $\;log|\,{\cal E}\,|\;$ & $\;log|\,c{\cal B}\,|\;$ &
    $\;log|\,{\cal E}\,|\;$ & $\;log|\,c{\cal B}\,|\;$ \\
\hline
\hline
0.0890 &-4.0004 &-14.6999  &-3.9998  &-14.6988  &-4.0009  &-14.6999  \\
\hline
0.1000 & -4.2028 &-5.8604  &-4.2021  &-5.8597  &-4.2036  &-5.8611  \\
\hline
0.2000 & -5.4069 &-6.0606  &-5.4040  &-6.0577  &-5.4098  &-6.0635  \\
\hline
0.3000 & -6.1112 &-6.4859  &-6.1048  &-6.4795  &-6.1178  &-6.4925  \\
\hline
0.5000 & -6.9986 &-7.0838  &-6.9807  &-7.0657  &-7.0166  &-7.1018  \\
\hline
0.8000 & -7.8147 &-7.6621  &-7.7690  &-7.6158  &-7.8607  &-7.708  \\
\hline
1.0000 & -8.2018 &-7.9419  &-8.1308  &-7.8695  &-8.2734  &-8.0138  \\
\hline
1.2000 & -8.5177 &-8.1723  &-8.4156  &-8.0678  &-8.6204  &-8.2753  \\
\hline
1.5000 & -8.9034 &-8.4557  &-8.7441  &-8.2913  & -9.0624 &-8.6154  \\
\hline
1.6000 & -9.0147 &-8.5378  &-8.8333  &-8.3505  &-9.1049  &-8.7191  \\
\hline
\end{tabular}
\end{center}
\end{table}

These results are very similar to the ones of our latter work.
Comparing the initial and the final values of the fields for each
geometry the data show an amplification of the ${\cal B}$ field of
the order $\approx 10^{+6}$, and an overall reduction of the
electric field of the order $\approx 10^{-5}$. It should be
noticed that these results are determined not only by the
evolution of the function $f(t)$ (which constrains the field
equations) but also by the direct contribution of the geometry as
given by the scale factor, $R(t)$, that is present in both
mathematical expressions for ${\cal E}$ and ${\cal B}$. It is the
interchange between the gravitational and electromagnetic fields
that imposes, as the Universe evolves, the decrease of the
electric field and the amplification of the ${\cal B}$ field.
\vspace{1cm}
{\raggedright
\section{Conclusion} }  \label{C}
\setcounter{equation}{0}

In all cases studied here, our results show the desired
amplification of the initial CMI, together with the reduction of
the electric field. The new version of the Dirac's \ae ther
incorporates in a natural way the 4-velocity (first pointed out by Dirac in 1951 
\cite{Dirac1951} which is interpreted as the velocity of the \ae ther
relative to an observer. This allow us to adapt our description to
any observer, inertial or not. In the case of a curved background,
an observer with $v_{\alpha} = R(t) \delta^{0}_{\alpha}$ will see
the same phenomenon of amplification of the magnetic induction and
reduction of the electric intensity that had already been observed
in \cite{Oliveira 2001}, in the context of a Proca electromagnetic field
in a Dirac \ae ther.

The geometric relations between the electromagnetic field and the
metric tensor involves the scale factor $R(t)$. The amplification
of the ${\cal B}$ field in this model is determined by the
coupling between gravitational and electromagnetic fields as in
the usual theory of electromagnetism in a curved background.
Larger couplings between the electromagnetic and the gravitational
fields could lead to an even faster and/or more intense
interchange, as can be seen in \cite{TW1988}. As the
{\it geometric amplification} describes how the expansion of the
universe influences the electromagnetic fields, there is also the
possibility that electromagnetic fields influence the expansion of
the universe \cite{MT2001}. Our results confirm once
again the strict relations between the electromagnetic and
gravitational phenomena.

The obtained geometric amplification is most probably superseded
by the conventional dynamo effect in the interior of stars, in the
intragalactic medium, and even in the scales of galaxies. However,
in scales of clusters and larger, where the dynamo action most
probably fails, the geometric amplification of the seed CMI may be
the only important effect to be considered. We believe that in the
near future technological advances will be able to detect these
fields and confirm our results.

\vspace{3mm}
{\raggedright
\section*{Acknowledgements}  } 
\label{Acknow}
\setcounter{equation}{0}
\begin{sloppypar}
All of us thank J. A. Helayel-Neto	(Centro Brasileiro de Pesquisas F\'{\i}sicas (CBPF)) for his friendly encouragement.
\end{sloppypar}

\end{document}